\documentclass[aps,prb,twocolumn,longbibliography,superscriptaddress,final]{revtex4-2}

\pdfoutput=1

\usepackage[colorlinks=true,allcolors=blue]{hyperref}
\usepackage{graphicx}
\usepackage{amsmath,amssymb,mathrsfs,array,longtable,delarray}
\usepackage{dcolumn}
\usepackage{bm}

\begin{document}


\title{Electroluminescence in dopant-free GaAs/AlGaAs single heterojunctions: \\ 2D free excitons, H-band, and the tidal effect} 

\author{N. Sherlekar}
\thanks{These authors contributed equally to this work}
\address{Institute for Quantum Computing, University of Waterloo, Waterloo N2L 3G1, Canada}
\address{Department of Electrical and Computer Engineering, University of Waterloo, Waterloo N2L 3G1, Canada}

\author{S. R. Harrigan}
\thanks{These authors contributed equally to this work}
\address{Institute for Quantum Computing, University of Waterloo, Waterloo N2L 3G1, Canada}
\address{Department of Physics and Astronomy, University of Waterloo, Waterloo N2L 3G1, Canada}
\address{Waterloo Institute for Nanotechnology, University of Waterloo, Waterloo N2L 3G1, Canada}

\author{L. Tian}
\thanks{These authors contributed equally to this work}
\address{Institute for Quantum Computing, University of Waterloo, Waterloo N2L 3G1, Canada}
\address{Department of Electrical and Computer Engineering, University of Waterloo, Waterloo N2L 3G1, Canada}

\author{B. Cunard}
\address{Institute for Quantum Computing, University of Waterloo, Waterloo N2L 3G1, Canada}
\address{Department of Electrical and Computer Engineering, University of Waterloo, Waterloo N2L 3G1, Canada}

\author{Y. Qi}
\address{Institute for Quantum Computing, University of Waterloo, Waterloo N2L 3G1, Canada}
\address{Department of Physics and Astronomy, University of Waterloo, Waterloo N2L 3G1, Canada}

\author{B. Khromets}
\address{Institute for Quantum Computing, University of Waterloo, Waterloo N2L 3G1, Canada}
\address{Department of Physics and Astronomy, University of Waterloo, Waterloo N2L 3G1, Canada}

\author{M. C. Tam}
\address{Department of Electrical and Computer Engineering, University of Waterloo, Waterloo N2L 3G1, Canada}
\address{Waterloo Institute for Nanotechnology, University of Waterloo, Waterloo N2L 3G1, Canada}

\author{\\H. S. Kim}
\address{Department of Electrical and Computer Engineering, University of Waterloo, Waterloo N2L 3G1, Canada}
\address{Waterloo Institute for Nanotechnology, University of Waterloo, Waterloo N2L 3G1, Canada}

\author{Z. R. Wasilewski}
\address{Institute for Quantum Computing, University of Waterloo, Waterloo N2L 3G1, Canada}
\address{Department of Electrical and Computer Engineering, University of Waterloo, Waterloo N2L 3G1, Canada}
\address{Department of Physics and Astronomy, University of Waterloo, Waterloo N2L 3G1, Canada}
\address{Waterloo Institute for Nanotechnology, University of Waterloo, Waterloo N2L 3G1, Canada}

\author{J. Baugh}
\email{baugh@uwaterloo.ca}
\address{Institute for Quantum Computing, University of Waterloo, Waterloo N2L 3G1, Canada}
\address{Department of Physics and Astronomy, University of Waterloo, Waterloo N2L 3G1, Canada}
\address{Waterloo Institute for Nanotechnology, University of Waterloo, Waterloo N2L 3G1, Canada}
\address{Department of Chemistry, University of Waterloo, Waterloo N2L 3G1, Canada}

\author{M. E. Reimer}
\email{mreimer@uwaterloo.ca}
\address{Institute for Quantum Computing, University of Waterloo, Waterloo N2L 3G1, Canada}
\address{Department of Electrical and Computer Engineering, University of Waterloo, Waterloo N2L 3G1, Canada}
\address{Department of Physics and Astronomy, University of Waterloo, Waterloo N2L 3G1, Canada}
\address{Waterloo Institute for Nanotechnology, University of Waterloo, Waterloo N2L 3G1, Canada}

\author{F. Sfigakis}
\email{francois.sfigakis@uwaterloo.ca}
\address{Institute for Quantum Computing, University of Waterloo, Waterloo N2L 3G1, Canada}
\address{Department of Electrical and Computer Engineering, University of Waterloo, Waterloo N2L 3G1, Canada}
\address{Department of Chemistry, University of Waterloo, Waterloo N2L 3G1, Canada}


\begin{abstract}
Bright electroluminescence (EL) from dopant-free ambipolar lateral p\nobreakdashes--n junctions in GaAs/AlGaAs single heterointerface (SH) heterostructures is used to probe neutral free excitons arising from two-dimensional electron and hole gases (2DEGs and 2DHGs). The EL spectra reveal both the heavy-hole neutral free exciton (X$^0$) and the high-energy free exciton of the H band (HE). A combination of transition energies, lifetimes, spatial emission profiles, and temperature dependences points to a predominantly two-dimensional character for these excitons at the SH. For X$^0$, the EL peak energies (1515.5–1515.7 meV) lie slightly above the corresponding bulk GaAs photoluminescence (PL) line at 1515.3 meV, while time-resolved measurements yield markedly shorter lifetimes for EL than for PL (337 ps vs. 1610 ps), consistent with recombination in a confined interfacial layer. The HE exciton exhibits a Stark blueshift under forward bias below threshold, and its energies and lifetimes (down to 575 ps) are tuned by the topgate voltage; above threshold, HE emission is quenched in favor of X$^0$. Finally, the tidal effect$-$a form of pulsed EL generated by swapping the topgate voltage polarity in ambipolar field-effect transistors$-$produces an X$^0$ line at the same energy as in the lateral p–n junction and reproduces the characteristic nonmonotonic frequency dependence of the brightness previously observed in quantum-well heterostructures, again indicating a 2D-like origin. Taken together, these results show electrically generated and controllable 2D-like excitons (HE and X$^0$), thereby bridging 2D exciton physics and 2DEG/2DHG platforms in dopant-free GaAs/AlGaAs SH devices.
\end{abstract}


\maketitle

Two-dimensional (2D) lateral p\nobreakdashes--n junctions have been realized in modulation-doped GaAs/AlGaAs heterostructures \cite{kaestnerNanoscaleLateralLight2003, cecchiniHighperformancePlanarLightemitting2003, cecchiniSurfaceAcousticWavedriven2004, hoseyLateralNpJunction2004, gellSurfaceacousticwavedrivenLuminescenceLateral2006, wunderlichExperimentalObservationSpinHall2005, lunghiAntibunchedPhotonsLateral2011, kaestnerQuasilateral2DEG2DHGJunction2003}, dopant-free GaAs/AlGaAs ambipolar field-effect transistors (FETs) \cite{daiLateralTwoDimensionalPin2013, daiHighqualityPlanarLight2014, chungQuantizedChargeTransport2019a, hsiaoSinglephotonEmissionSingleelectron2020, tianStableElectroluminescenceAmbipolar2023}, and undoped semiconductor 2D material FETs \cite{pospischilSolarenergyConversionLight2014, baugherOptoelectronicDevicesBased2014, rossElectricallyTunableExcitonic2014}. Unlike conventional III-V semiconductor light emitting diodes, lateral p\nobreakdashes--n junctions consist of a 2D electron gas (2DEG) and a 2D hole gas (2DHG) contiguous to each other in the same plane. Due to the wide-ranging functionalities of their 2DEG/2DHG and their ease of integration/scalability with other semiconductor optolelectronic devices, lateral p\nobreakdashes--n junctions are a promising platform for photonic applications, such as spin-photon conversion \cite{hsiaoSinglephotonEmissionSingleelectron2020, oiwaConversionSinglePhoton2017, gaudreauEntanglementDistributionSchemes2017, fujitaAngularMomentumTransfer2019} and single photon sources \cite{hsiaoSinglephotonEmissionSingleelectron2020, blumenthalGigahertzQuantizedCharge2007, buonacorsiNonadiabaticSingleelectronPumps2021}.

\begin{figure*}[t]
    \includegraphics[width=2.0\columnwidth]{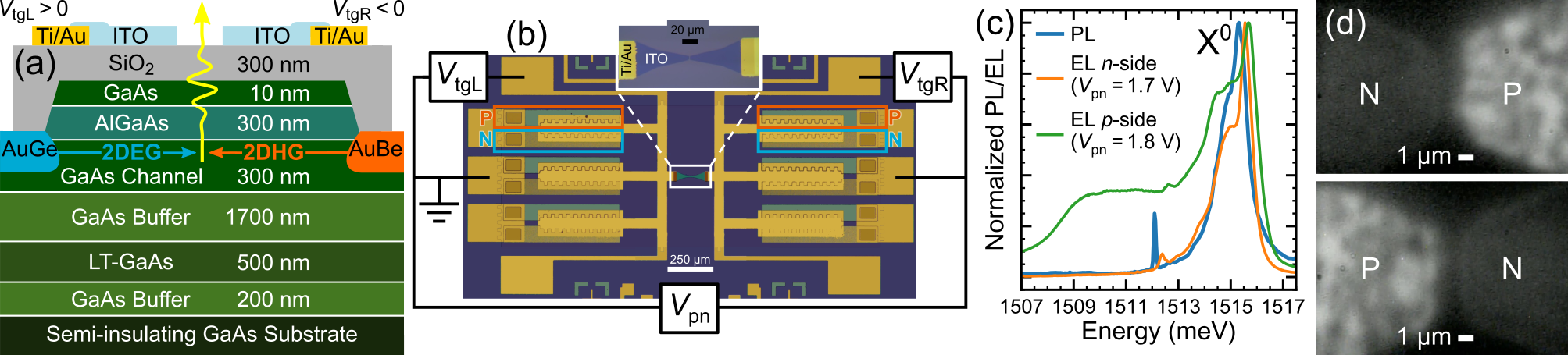}
    \caption{Sample A. (a) Cross-sectional schematic of a lateral p\nobreakdashes–n junction in a dopant-free FET with a GaAs/AlGaAs SH heterostructure. (b) Microscope image of fabricated device with associated circuit. $V_{\text{tgL}}$ and $V_{\text{tgR}}$ are the voltages applied to the left and right topgates respectively. $V_{\text{pn}}$ is the forward bias (or source-drain bias). The device is ambipolar, i.e. there are $p$-type and $n$-type phmic contacts on both sides of the p\nobreakdashes--n junction. (c) EL spectra from the $n$-side and $p$-side of the device superimposed onto the PL spectrum from the same heterostructure wafer. The X$^0$ peak is visible in all three spectra, all within a 0.4 meV window. Otherwise, the $p$-side EL is markedly different from PL and the $n$-side EL. (d) EL intensity maps of the device in two measurement configurations (PN and NP), illustrating that the $p$-side is always brighter than the $n$-side.}
    \label{Fig:Fig1}
\end{figure*}

In this Letter, we present bright EL from lateral p\nobreakdash–n junctions and ambipolar FETs, both fabricated in dopant-free GaAs/AlGaAs single heterointerface (SH) heterostructures. We demonstrate two types of true 2D free excitons, each generated within a 2DEG or 2DHG, in these devices. All previous reports of GaAs/AlGaAs lateral p\nobreakdashes--n junctions used quantum well (QW) heterostructures \cite{kaestnerNanoscaleLateralLight2003, cecchiniHighperformancePlanarLightemitting2003, cecchiniSurfaceAcousticWavedriven2004, hoseyLateralNpJunction2004, gellSurfaceacousticwavedrivenLuminescenceLateral2006, wunderlichExperimentalObservationSpinHall2005, lunghiAntibunchedPhotonsLateral2011, kaestnerQuasilateral2DEG2DHGJunction2003, daiLateralTwoDimensionalPin2013, daiHighqualityPlanarLight2014, chungQuantizedChargeTransport2019a, hsiaoSinglephotonEmissionSingleelectron2020,tianStableElectroluminescenceAmbipolar2023}.  All previously observed X$^0$ (whether from photoluminescence or electroluminescence) in SH heterostructures (with doping or not) were 3D excitons.

We use two different methods for generating EL, which both show 2D free excitons. The first method involves ambipolar lateral p\nobreakdashes--n junctions, and two types of 2D free excitons are observed. The first type is the high-energy (HE) exciton of the H\nobreakdash-band \cite{yuanNewPhotoluminescenceEffects1985, zhaoRadiativeRecombinationDoped1990, bergmanTimeresolvedMeasurementsRadiative1991, shenPhotoluminescenceModulationdoped1999, kubisaPhotoluminescenceInvestigationsTwodimensional2003, quOpticalPropertiesHband2003, quHbandEmissionSingle2003, kundrotasEnhancedExcitonPhotoluminescence2010, kenEffectElectricCurrent2020}, previously only observed in photoluminescence (PL). HE is known to be a spatially-indirect 2D free exciton \cite{kubisaPhotoluminescenceInvestigationsTwodimensional2003, kundrotasEnhancedExcitonPhotoluminescence2010, kenEffectElectricCurrent2020}. The second type is the heavy-hole neutral free exciton (X$^0$). Comparing the contrasting energies and lifetimes of X$^0$ from EL and PL in the same SH heterostructure strongly suggest that the nature of EL X$^0$ is truly 2D whereas that of PL X$^0$ is definitively 3D (bulk). The second method for generating EL is the tidal effect from ambipolar FETs, a form of pulsed EL typically reported in undoped 2D materials \cite{lienLargeareaBrightPulsed2018, paurElectroluminescenceMultiparticleExciton2019, zhaoGenericElectroluminescentDevice2020, zhuHighEfficiencyWavelengthTunableMonolayer2021} but also recently observed in GaAs/AlGaAs QW heterostructures \cite{harriganPulsedElectroluminescenceDopantfree2025}. Its EL X$^0$ energy matches the EL X$^0$ from our lateral p\nobreakdashes--n junctions. The tidal effect in SH heterostructures is only approximately three times less bright than in QW heterostructures. These two observations strongly suggest a 2D origin for its excitons.

\textbf{\textit{Methods and initial characterization}}. Figure~\ref{Fig:Fig1}(a) shows a cross-sectional schematic of a typical device fabricated in a SH heterostructure; see section I.A of the supplemental material for growth and fabrication details \cite{supplementalPRB-SH-2DPN}. For comparison, we also used lateral p\nobreakdashes--n junctions fabricated in QW heterostructures; these are described elsewhere \cite{tianStableElectroluminescenceAmbipolar2023, harriganPulsedElectroluminescenceDopantfree2025}. All our devices are ambipolar: $p$\nobreakdash-type and $n$\nobreakdash-type ohmic contacts are present on both sides of the p\nobreakdashes--n junction. Voltages are applied to the topgates on the left ($V_{\text{tgL}}$) and right ($V_{\text{tgR}}$) of the junction, inducing a 2DEG and 2DHG on opposite sides. Figure~\ref{Fig:Fig1}(b) shows an optical image of a typical device, as well as the electrical circuit used in experiments. The diode turn-on at the applied source-drain bias $V_{\text{pn}}$~$\approx$~1.52~V matches the bandgap of GaAs ($E_{\text{g}}$~$\approx$~1.52~eV). To overcome the problem of the rapid quenching of EL during standard DC operation, the device was operated with the set-reset voltage sequence \cite{tianStableElectroluminescenceAmbipolar2023} (see section I.B of the supplemental material \cite{supplementalPRB-SH-2DPN}). Unless otherwise noted, all data shown here was collected from two nominally identical SH devices (samples A and B) at a temperature of $T$~=~1.6~K. Maximum mobilities for our SH heterostructure are $7.3 \times 10^6$~cm$^2$\,V$^{-1}$\,s$^{-1}$ at an electron density of $2.6 \times 10^{11}$~cm$^{-2}$ ($V_{\text{tg}}$~=~$+$\,5~V) for the 2DEG and $6.2 \times 10^5$~cm$^2$\,V$^{-1}$\,s$^{-1}$ at a hole density $2.0 \times 10^{11}$~cm$^{-2}$ ($V_{\text{tg}}$~=~$-$\,5~V) for the 2DHG, obtained from a dedicated ambipolar reference Hall bar (see Fig.~S5 of the supplemental material \cite{supplementalPRB-SH-2DPN}).

Figure~\ref{Fig:Fig1}(c) shows normalized EL spectra captured from the $n$\nobreakdash-side of the junction and the $p$\nobreakdash-side of the junction superimposed onto the normalized PL spectrum. We identify the most prominent peak in all three spectra as X$^0$. The two EL X$^0$ energies (1515.5 meV and 1515.7 meV) are slightly larger than but close to the bulk GaAs PL X$^0$ energy (1515.3 meV) \cite{pavesiPhotoluminescenceAlxGa1xAsAlloys1994, schusterElectricFieldDistribution2016}. The $p$\nobreakdash-side EL differs from PL and the $n$\nobreakdash-side EL, due to the presence of an additional broad emission band below 1512 meV. This broad emission band is primarily responsible for the brighter integrated EL intensity on the $p$\nobreakdash-side relative to the $n$\nobreakdash-side in camera images of the p\nobreakdashes--n junction [see Fig.~\ref{Fig:Fig1}(d)].

\begin{figure}[t]
    \includegraphics[width=1.0\columnwidth]{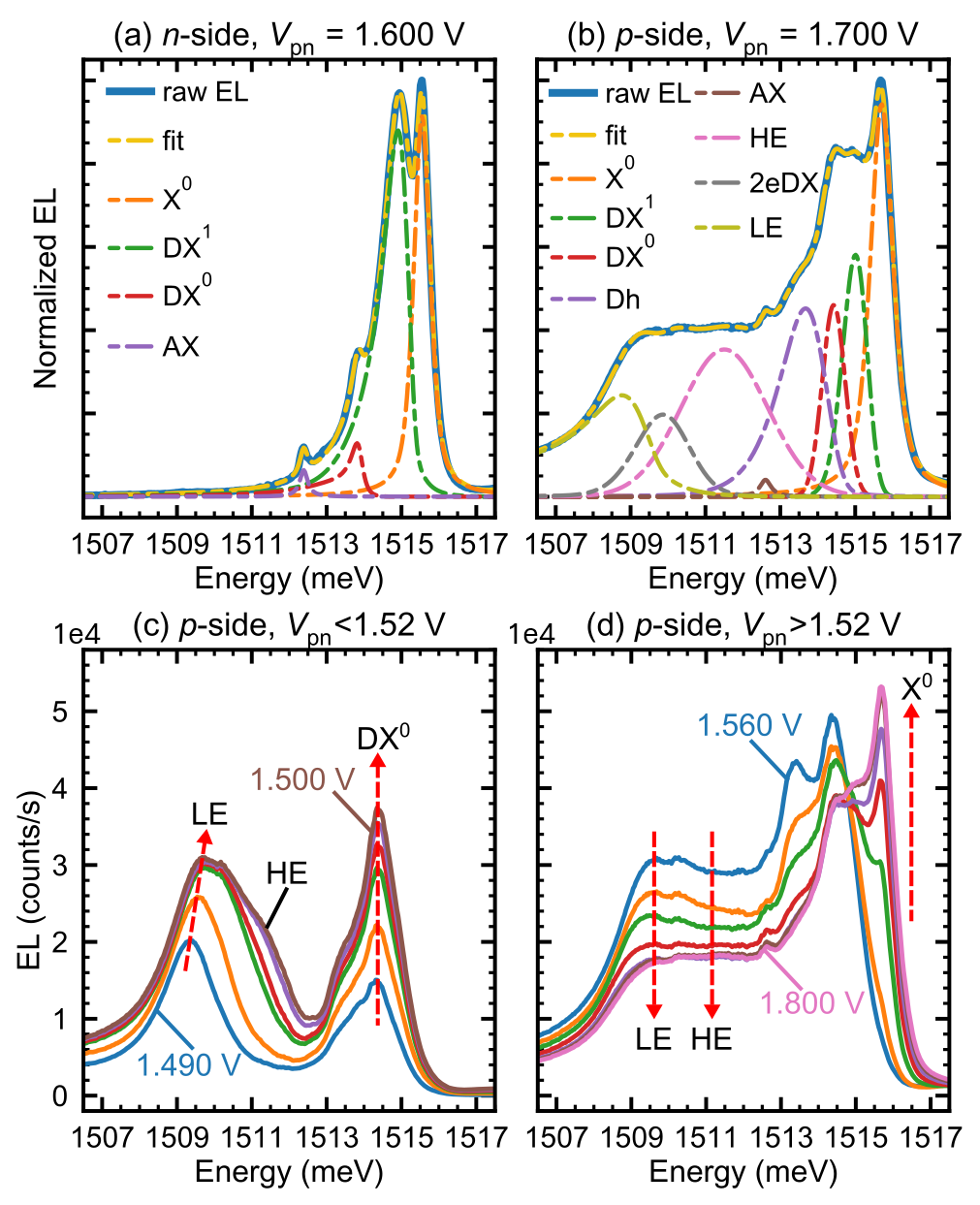}
    \caption{EL from a lateral p–n junction at a SH interface (sample A). Normalized spectra with peak fits for: (a) the $n$-side and (b) the $p$-side. Corresponding exciton type and energies are listed in Table~S2 of the supplemental material \cite{supplementalPRB-SH-2DPN}. Spectra from the $p$-side as a function of $V_{\text{pn}}$ for: (c) subthreshold operation and (d) above-threshold operation. For (c), $V_{\text{pn}}$ is increased from 1.490\,V to 1.500\,V in steps of 2\,mV. For (d), $V_{\text{pn}}$ is increased from 1.560\,V to 1.800\,V in steps of 40\,mV.}
    \label{Fig:Fig2}
\end{figure}

\textbf{\textsl{2D} \textit{HE in lateral p\nobreakdashes--n junctions}}. Constituent lines of the EL spectra (with energy $E_{\textsc{el}}$) from the $n$\nobreakdash-side and $p$\nobreakdash-side were identified by fitting them to Pseudo-Voigt lineshapes \cite{stancikSimpleAsymmetricLineshape2008}, as seen in Figs.~\ref{Fig:Fig2}(a) and \ref{Fig:Fig2}(b). All emission lines observed on the $n$\nobreakdash-side are also visible on the $p$\nobreakdash-side. Most  peaks and peak shoulders in both spectra result from known \cite{pavesiPhotoluminescenceAlxGa1xAsAlloys1994} donor (D) and acceptor (A) impurities (see Table~S2 and Fig.~S6 in the supplemental material \cite{supplementalPRB-SH-2DPN}). We identify the broad emission band below 1512 meV on the $p$\nobreakdash-side as the H\nobreakdash-band \cite{yuanNewPhotoluminescenceEffects1985, zhaoRadiativeRecombinationDoped1990, bergmanTimeresolvedMeasurementsRadiative1991, shenPhotoluminescenceModulationdoped1999, kubisaPhotoluminescenceInvestigationsTwodimensional2003, quOpticalPropertiesHband2003, quHbandEmissionSingle2003, kundrotasEnhancedExcitonPhotoluminescence2010, kenEffectElectricCurrent2020}, which arises from the formation of an energy ‘dimple’ in the conduction band underneath and in close proximity to the 2DHG (see Fig.~6 in Ref.~\onlinecite{kenEffectElectricCurrent2020}). The dimple is induced by the 2DHG itself, which also screens the dimple from the topgate. Some of the electrons crossing the junction from the $n$\nobreakdash-side to the $p$\nobreakdash-side fall into this potential dimple, becoming confined to a plane parallel to the 2DHG. Two distinct transition lines make up the H\nobreakdash-band \cite{kenEffectElectricCurrent2020}. The symmetric HE line corresponds to the recombination of a free hole from the 2DHG and an electron that freely moves in the plane of the conduction band dimple. Thus HE is a true 2D free exciton. The highly asymmetric low-energy (LE) line (with a long low-energy tail) is due to the recombination of a free hole from the 2DHG and an electron from the dimple that has been captured by a nearby donor impurity. Both HE and LE disappear at $T$~$\approx$~16~K (see Fig.~S7 in supplemental material \cite{supplementalPRB-SH-2DPN}), matching previously reported behavior in PL \cite{yuanNewPhotoluminescenceEffects1985}.

Figures~\ref{Fig:Fig2}(c) and \ref{Fig:Fig2}(d) show contrasting trends of the $p$\nobreakdash-side EL spectra as a function of increasing $V_{\text{pn}}$ for subthreshold operation ($V_{\text{pn}}<1.52$~V) and above-threshold operation ($V_{\text{pn}}>1.52$~V), respectively. Topgates on both sides of the p\nobreakdashes--n junction are fixed at $|V_{\text{tg}}|$ = 5.0~V.

With increasing $V_{\text{pn}}$ during subthreshold operation [Fig.~\ref{Fig:Fig2}(c)], three trends are observed:\\
\indent (1) the integrated intensity of the H\nobreakdash-band increases,\\
\indent (2) the intensity of HE increases faster than LE, and\\
\indent (3) both LE and HE are blueshifted (Stark effect).\\
Trend (1) occurs due to higher current through the device. Trends (2) and (3) were reported for PL H\nobreakdash-band, when decreasing the magnitude of an applied in-plane electric field [see Fig.~4(a) in Ref.~\onlinecite{kenEffectElectricCurrent2020}]. For any p\nobreakdash–n junction in general, increasing $V_{\text{pn}}$ during subthreshold operation effectively decreases the built-in potential. This in turn reduces the associated in-plane electric field across a lateral p\nobreakdash–n junction, explaining trends (2) and (3) in our devices.

With increasing $V_{\text{pn}}$ during above-threshold operation [Fig.~\ref{Fig:Fig2}(d)], the HE and LE emission energies remain constant, since the in-plane electric field across the p\nobreakdashes--n junction no longer changes significantly. For $V_{\text{pn}}>1.56$~V, the intensities of HE, LE, and bound excitons (e.g., Dh and DX$^0$) decrease as a function of $V_{\text{pn}}$ in favor of the higher energy transition X$^0$.

\begin{figure}[t]
    \includegraphics[width=1.0\columnwidth]{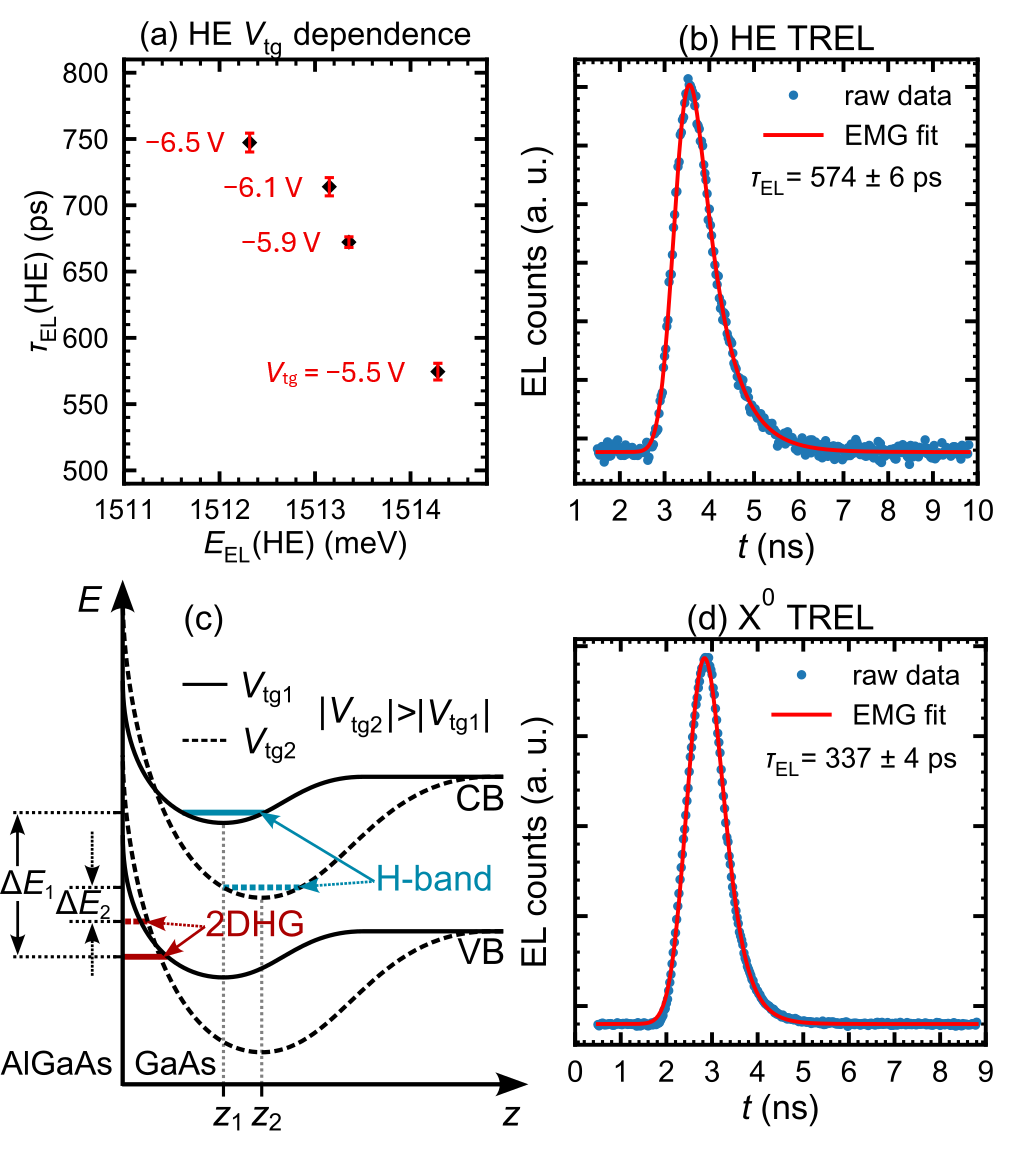}
    \caption{Sample A. (a) HE exciton lifetimes at different energies. (b) Fit (red solid line) to TREL experiments (blue circles) at the highest HE energy, using an exponentially-modified Gaussian (EMG). (c) Schematic of band structure, showing changes in the conduction band (CB) and the valence band (VB) at $V_{\text{tg1}}$ and $V_{\text{tg2}}$ on the $p$-side, with $\Delta E_1$ and $\Delta E_2$ corresponding to different $E_{\textsc{el}}$(HE). $z_1$ and $z_2$ are the corresponding spatial separations between the recombining electron and hole. (d) TREL and EMG fit for EL X$^0$ from the $p$-side. }
    \label{Fig:Fig3}
\end{figure}

By adjusting $V_{\text{tg}}$ on the $p$\nobreakdash-side, Figure~\ref{Fig:Fig3}(a) demonstrates HE can be tuned to different peak energies $E_{\textsc{el}}$ (see also Fig.~S8 in the supplemental material \cite{supplementalPRB-SH-2DPN}). Time-resolved EL (TREL) measurements were conducted at these energies, and the corresponding lifetimes $\tau_{\textsc{el}}$ were obtained by fitting the data to an exponentially-modified Gaussian (EMG) [e.g., Fig.~\ref{Fig:Fig3}(b)]. Figure~\ref{Fig:Fig3}(a) reveals a decrease of $\tau_{\textsc{el}}$ with increasing $E_{\textsc{el}}$. This trend can be explained by the simplified band structure schematic in Fig.~\ref{Fig:Fig3}(c), where the H\nobreakdash-band dimple is depicted for two different $V_{\text{tg}}$ (where $|V_{\text{tg2}}|$~$>$~$|V_{\text{tg1}}|$). The 2DHG density is higher for $V_{\text{tg2}}$, with a deeper H\nobreakdash-band dimple and a lower $E_{\textsc{el}}$(HE) since $\Delta E_2 < \Delta E_1$. Simultaneously, the distance between the H\nobreakdash-band dimple and the GaAs/AlGaAs interface increases ($z_2 > z_1$). As a result, the electron-hole wavefunction overlap of the HE exciton reduces, which in turn increases $\tau_{\textsc{el}}$(HE).

\begin{figure}[t]
    \includegraphics[width=1.0\columnwidth]{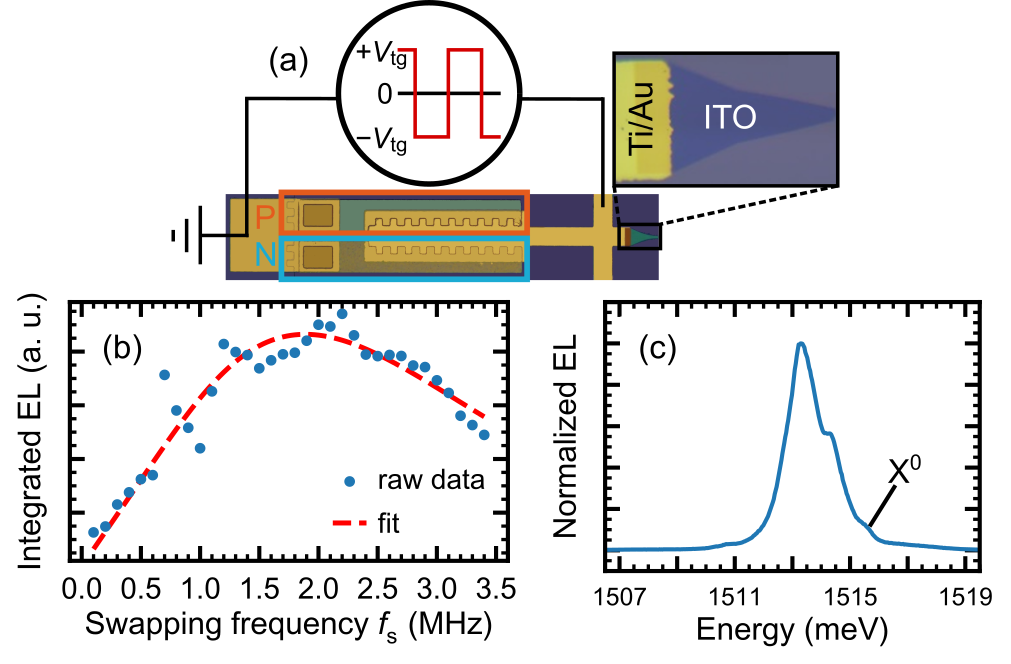}
    \caption{EL from the tidal effect in a SH heterostructure (sample B). (a) Electrical circuit. (b) Total EL intensity as a function of swapping frequency $f_{\text{s}}$ with $V_{\text{tg}}$\,=\,5\,V. The dashed line is a fit to Eq.\,(1). (c) Tidal effect spectrum, using $f_{\text{s}}$\,=\,800\,kHz and $V_{\text{tg}}$\,=\,5\,V. }
    \label{Fig:Fig4}
\end{figure}

\textbf{\textsl{2D} \textit{X$^0$ in lateral p\nobreakdashes--n junctions}}. For both SH and QW heterostructures, almost all EL is generated directly below the topgates near the edge of the p\nobreakdashes--n junction, not in the gap between the $p$-side and $n$-side topgates [e.g., see Fig.~\ref{Fig:Fig1}(d)]. Furthermore, the EL brightness from both heterostructure types are similar to within an order of magnitude, with the QW always being brighter than the SH heterostructure. Radiative recombination (whether from the $p$-side or $n$-side) must occur at the GaAs/AlGaAs interface, since all majority carriers are confined at that location. The slightly larger EL X$^0$ energies than the PL X$^0$ energy are consistent with this interpretation.

In the SH heterostructure from Fig.~\ref{Fig:Fig1}(a), the X$^0$ lifetimes are $\tau_{\textsc{el}}$\,=\,337\,ps for EL from a p\nobreakdashes--n junction [see Fig.~\ref{Fig:Fig3}(d)] and $\tau_{\textsc{pl}}$\,=\,1610\,ps for PL (see Fig.~S9 in the supplemental material \cite{supplementalPRB-SH-2DPN}). In our QW heterostructure, the same measurements yielded $\tau_{\textsc{el}}$\,=\,237\,ps and $\tau_{\textsc{pl}}$\,=\,419\,ps, respectively [see Fig.~4(b) from Ref.~\onlinecite{tianStableElectroluminescenceAmbipolar2023}]. Both EL X$^0$ and PL X$^0$ consist of 2D excitons confined to the QW plane, so their values are similar in range. The ratio $\tau_{\textsc{pl}}/\tau_{\textsc{el}}=1.8$ for our QW represents the extra scattering present in EL relative to PL \cite{tianStableElectroluminescenceAmbipolar2023}. In our SH heterostructure, we speculate that the ratio $\tau_{\textsc{pl}}/\tau_{\textsc{el}}=4.8$ is much larger because EL involves 2D excitons whereas PL involves 3D excitons from the 2 \textmu m thick GaAs layer (0.3 \textmu m channel $+$ 1.7 \textmu m buffer). Similar results were obtained for EL and PL in different SH and QW heterostructures (see Table S4 and Figs.~S9$-$S10 in the supplemental material \cite{supplementalPRB-SH-2DPN}).

At higher temperatures, EL X$^0$ disappears above $T$\,$\approx$\,24\,K (see Fig.~S7 in supplemental material \cite{supplementalPRB-SH-2DPN}), whereas HE disappears at $T$\,$\approx$\,16\,K. The higher operating temperature of EL X$^0$ indicates it has higher binding and bound state energies than HE, which suggests EL X$^0$ is a 2D exciton.

Without a bottom barrier, why aren't minority carriers repelled into the substrate before recombining with majority carriers? The answer lies with external electric field profiles, which we break down into their in-plane and out-of-plane components with respect to the 2DEG/2DHG plane. The in-plane electric field at the center of the gap region between the 2DEG and 2DHG will be approximately $E_\parallel$~$\approx$~$V_\text{pn}$/gap~=~1.5\,V/5\,\textmu m~$=$~3~kV/cm. From symmetry considerations, the out-of-plane electric field must vanish $E_\perp \approx 0$ at the center of the gap (see Fig.~S11 and associated discussion in the supplemental material \cite{supplementalPRB-SH-2DPN}). Electrons from the 2DEG are rapidly accelerated across the gap (transit time\,$\approx$\,20$-$70\,ps; see Table~S5 and associated discussion in the supplemental material \cite{supplementalPRB-SH-2DPN}) along the GaAs/AlGaAs interface, spending little time interacting with holes traveling in the opposite direction. Thus negligible EL [20$-$70\,ps~$\ll$~$\tau_\textsc{el}$~$\approx$~340\,ps from Fig.~\ref{Fig:Fig3}(d)] is generated in the low-density gap region, as can be seen in Fig.~\ref{Fig:Fig1}(d). Accelerated carriers from one side are then injected directly into the plane of the 2DEG/2DHG underneath the topgate from the opposite side, where they are drastically slowed down by attractive interactions with surrounding majority carriers. The latter also provide intra-layer screening \cite{andoElectronicProperties2DSystems1982} to the minority carriers from the topgate, momentarily preventing their expulsion into the substrate and significantly increasing their dwell time near majority carriers. This in turn enables significant EL to be generated underneath the topgate, as can be seen in Fig.~\ref{Fig:Fig1}(d).

\textbf{\textit{The tidal effect}}. We demonstrate the so-called tidal effect \cite{lienLargeareaBrightPulsed2018, paurElectroluminescenceMultiparticleExciton2019, zhaoGenericElectroluminescentDevice2020, zhuHighEfficiencyWavelengthTunableMonolayer2021, harriganPulsedElectroluminescenceDopantfree2025}, a form of pulsed EL, with the electrical circuit shown in Fig.~\ref{Fig:Fig4}(a). A periodic square-wave voltage (with frequency $f_{\text{s}}$ and peak-to-peak amplitude $2V_{\text{tg}}$) is applied to only one topgate of the device, while all ohmic contacts and the other topgate remain grounded during operation. The part of the device with the grounded topgate is not involved during the following. Swapping voltages between $+V_\text{tg}$ and $-V_\text{tg}$ causes carriers underneath the topgate to alternate between a 2DEG and 2DHG. During each topgate voltage swap, electrons (holes) previously induced under the topgate recede to the ohmic contacts, while simultaneously new holes (electrons) flow in from the ohmic contacts. Some of the electrons and holes moving in opposite directions radiatively recombine, generating EL pulses everywhere underneath the topgate. Radiative recombination must mostly occur at the GaAs/AlGaAs interface, since all incoming majority carriers are confined there.

A distinctive feature of the tidal effect is its nonmonotonic dependence of the integrated intensity~$I_\Sigma$ on swapping frequency~$f_{\text{s}}$, as shown in Fig.~\ref{Fig:Fig4}(b). This behavior can be modeled by \cite{harriganPulsedElectroluminescenceDopantfree2025}:
\begin{equation}
    I_\Sigma(f_{\text{s}}) = \frac{c f_{\text{s}} N_0}{2}~\text{erfc}\!\left(\frac{xf_{\text{s}}-\mu E}{2\sqrt{Df_{\text{s}}}}\right),
    \label{eqn1}
\end{equation}
\noindent where $c$ is a scaling factor accounting for recombination and collection efficiencies, $N_0$ is the steady-state carrier density, $x$ is the distance charge carriers have to travel between the ohmic contacts and the location where EL is collected, $\mu$ is the carrier mobility, $E$ is the in-plane electric field, and $D$ is the diffusion constant. In our devices, $x$~=~375~\textmu m, $E$~$\approx$~40~V\,cm$^{-1}$ (obtained from $E_{\text{g}}/ex$, where $e$ is the electron charge), and $N_0 = 2.0\times 10^{11}$~cm$^{-2}$ (the lesser of the 2DEG and 2DHG densities at $|V_\text{tg}|$~=~5~V). Using $\mu$ and $D$ as free parameters in Eq.\,(\ref{eqn1}), we fit the data in Fig.~\ref{Fig:Fig4}(c), and obtain $\mu = (2.3 \pm 0.1) \times 10^3$~cm$^2$\,V$^{-1}$\,s$^{-1}$ and $D = (4.0 \pm 0.7) \times 10^2$~cm$^2$\,s$^{-1}$. These values are roughly three orders of magnitude lower than corresponding values from our reference Hall bar, but close to the values in bulk GaAs \cite{ruchTransportPropertiesGaAs1968}. They are also very close to the values obtained for the tidal effect in GaAs QWs \cite{harriganPulsedElectroluminescenceDopantfree2025}. For both SH and QW heterostructures, Thomas-Fermi screening \cite{andoElectronicProperties2DSystems1982,shettyEffectsBiasedUnbiased2022} (the principal mechanism responsible for the high mobility in our reference Hall bar) is almost nonexistent at the wavefront of the incoming carriers, and therefore $\mu$ and $D$ are close to that of bulk GaAs during the tidal effect.

Figure~\ref{Fig:Fig4}(c) shows a tidal effect spectrum with a weak X$^0$ at 1515.5~meV, matching the $n$\nobreakdash-side EL X$^0$. Using a spectrometer with a cryogenic objective lens in back-to-back sample cooldowns to ensure identical optical setups, we observed the tidal effect to be three times brighter in the QW heterostructure than in the SH heterostructure. This small relative difference observed between our SH and QW devices implies significant electron-hole wavefunction overlap in our SH heterostructures, which is only possible if radiative recombination occurs at the GaAs/AlGaAs interface.

\textbf{\textit{Conclusions}}. We demonstrated bright EL in dopant-free ambipolar FETs and lateral p\nobreakdashes--n junctions fabricated from GaAs/AlGaAs SH heterostructures. We observed the H\nobreakdash-band's HE in EL, whose energy and lifetime can be tuned by the topgate voltage
in the above-threshold regime ($eV_\text{pn}>E_\text{gap}$). In the subthreshold regime ($eV_\text{pn}<E_\text{gap}$), the HE energy can also be tuned by the forward bias. We observed EL X$^0$ from lateral p\nobreakdashes--n junctions in a SH heterostructure: its lifetime, energy, emission location, and operating temperature all suggest a 2D origin, unlike PL X$^0$ observed in the same SH heterostructure with a 3D origin. Finally, we observed the tidal effect in a semiconductor SH heterostructure. The presence of EL X$^0$ in its spectra and a broadly similar brightness as QW heterostructures suggest its excitons may have a 2D origin.

\section*{Acknowledgments}

The authors thank members of the Quantum Photonic Devices Lab and Christine Nicoll for valuable discussions. This research was undertaken thanks in part to funding from the Canada First Research Excellence Fund (Transformative Quantum Technologies), Defence Research and Development Canada (DRDC), and Canada's Natural Sciences and Engineering Research Council (NSERC). S.R.H. acknowledges further support from the NSERC Canada Graduate Scholarships $-$ Doctoral program. The University of Waterloo's QNFCF facility was used for this work. This infrastructure would not be possible without the significant contributions of CFREF-TQT, Canada Foundation for Innovation (CFI), Innovations, Science and Economic Development Canada (ISED), the Ontario Ministry of Research, Innovation and Science, and Mike and Ophelia Lazaridis. Their support is gratefully acknowledged.

\section*{Data availability}

The data that support the findings of this study are available from the corresponding author upon reasonable request.

\end{document}


\title{{\LARGE{SUPPLEMENTARY MATERIAL}} \\~\\~\\
Electroluminescence in GaAs/AlGaAs single heterojunctions: \\
2D free excitons, H-band, and the tidal effect }

\author{~\\ N. Sherlekar, S. R. Harrigan, L. Tian, B. Cunard, Y. Qi, B. Khromets, \\ M. C. Tam,
H. S. Kim, Z. R. Wasilewski, J. Baugh, M. E. Reimer, and F. Sfigakis}

\affiliation{University of Waterloo, Waterloo N2L 3G1, Canada}

\maketitle

~\\~\\ \qquad\qquad \textbf{Table of Contents:}\\~\vspace{-5mm}\\
\indent\qquad\qquad Section \ref{sec:methods}: Experimental methods \\
\indent\qquad\qquad Section \ref{sec:transport}: Transport characterization, mobility and density\\
\indent\qquad\qquad Section \ref{sec:identification}: EL peak identification \\
\indent\qquad\qquad Section \ref{sec:Tdep}: H-band and X$^0$ temperature dependence \\
\indent\qquad\qquad Section \ref{sec:TREL-Hband}: H-band time-resolved EL \\
\indent\qquad\qquad Section \ref{sec:TRELPL-X0}: X$^0$ time-resolved EL and PL \\
\indent\qquad\qquad Section \ref{sec:discussion}: Electric field profiles and transit times


\clearpage
\newpage

\section{Experimental methods}
\label{sec:methods}

\subsection{Wafer growth \& sample fabrication}
\label{sec:mbefab}

\begin{figure}[h]
    \centering
    \includegraphics[width=0.35\columnwidth]{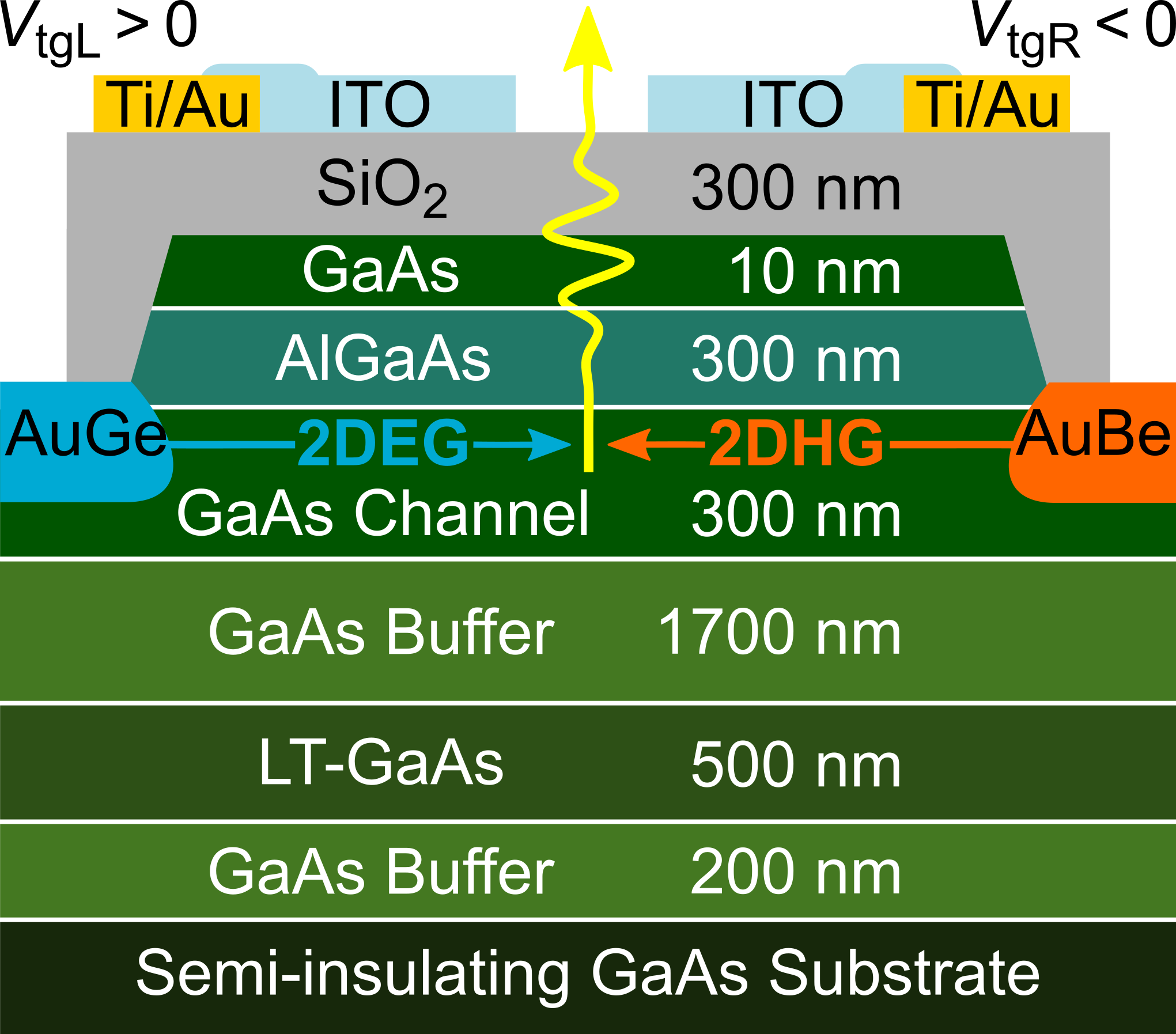}
    \caption{Cross-sectional schematic of the lateral p\nobreakdashes--n junction in wafer G265, a dopant-free GaAs/AlGaAs heterostructure with a single heterointerface (SH). Reproduction of Fig.\,1(a) from the main text.}
    \label{FigS1}
\end{figure}

The lateral p\nobreakdashes--n junction was made using a GaAs/AlGaAs heterostructure wafer (labeled G265) grown using molecular beam epitaxy (MBE), with carriers flowing along the high-mobility crystal direction $[1\bar{1}0]$. No intentional dopants were added to the wafer. The details of each layer in the MBE wafer stack (starting from the semi-insulating GaAs substrate) are as follows (see Fig.\,\ref{FigS1}):~200\,nm GaAs buffer, 500\,nm LT-GaAs layer, 1700\,nm GaAs buffer, 300\,nm GaAs channel layer, 300\,nm Al$_{0.3}$Ga$_{0.7}$As top barrier, 10\,nm GaAs cap layer. 

The fabrication steps for the lateral p\nobreakdashes--n junction are similar to those described for unipolar and ambipolar FETs.\cite{mak_distinguishing_2010, chenFabricationCharacterizationAmbipolar2012, tanejaNtypeOhmicContacts2016} After defining a mesa using a wet etch process, Ni/AuGe/Ni (\textit{n}-type) and AuBe (\textit{p}-type) recessed ohmic contacts were deposited and annealed at 450\,\textdegree{}C for 180\,s and 520\,\textdegree{}C for 180\,s, respectively. Next, 300\,nm thick PECVD SiO$_2$ was deposited over the device as an insulating layer. Above this, Ti/Au was deposited to form the inducing topgates. Close to the p\nobreakdashes--n junction, a 30\,nm semi-transparent conductive ITO layer is used as the topgate material to allow emitted light to pass through. A topgate bias $V_\text{tg}$ of sufficient magnitude induces a two-dimensional electron/hole gas (2DEG/2DHG) at the single heterointerface (SH) between the GaAs channel layer and the AlGaAs top barrier. The gap separating the left and right topgates is $\sim$ 5 \textmu{}m wide.

\subsection{Electrical and optical setup}
\label{sec:setups}

Devices were placed in an attocube attoDRY2100 closed-cycle optical cryostat, and cooled to $\sim$1.6\,K. Electrical connections were made using in-house custom electrical feedthroughs paired with existing cabling. DC voltages were applied to the topgates using SRS SIM928 isolated voltage sources, and a Keithley 2401 SourceMeter was used to apply the p\nobreakdashes--n junction source-drain bias ($V_\text{pn}$) across the ohmic contacts.

\begin{figure}[b]
    \includegraphics[width=1.0\columnwidth]{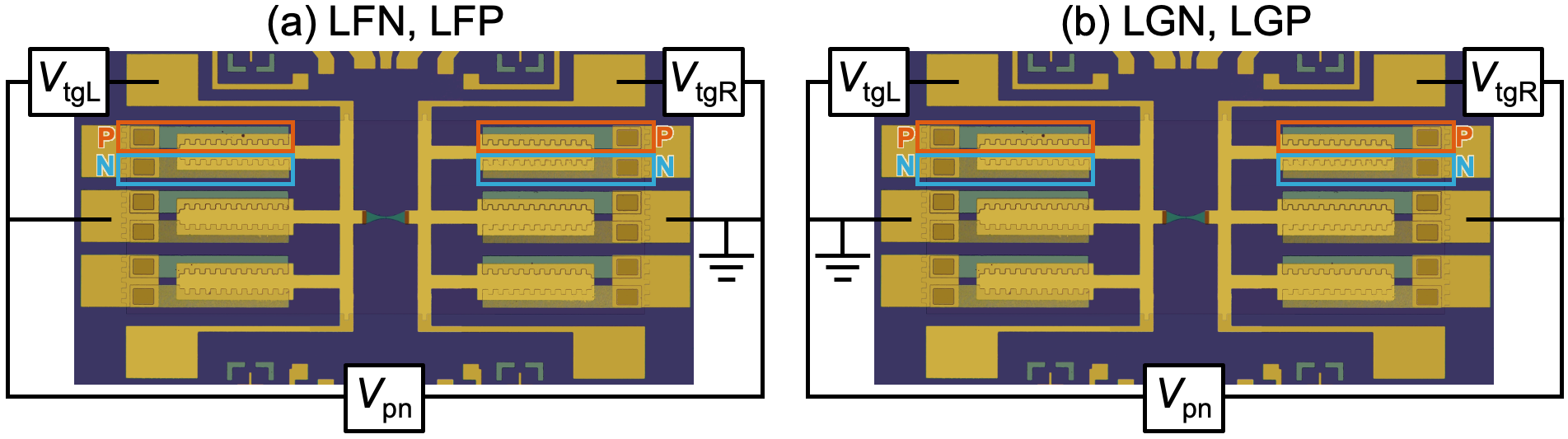}
    \caption{Circuit diagrams showing the four distinct configurations our ambipolar lateral p\nobreakdashes--n junctions can be connected in. $V_\text{tgL}$ and $V_\text{tgR}$ are the voltages applied on the left and right topgates respectively. $V_\text{pn}$ is the forward bias applied to the device. (a)~LFN or LFP configuration. The ohmic contacts on the left (\textit{n}-type or \textit{p}-type) are floating, and those on the right (\textit{p}-type or \textit{n}-type, respectively) are grounded. (b)~LGN or LGP configuration. The ohmic contacts on the left are grounded, and those on the right are floating.}
    \label{FigS2}
\end{figure}

The set-reset voltage sequence (as per Tian et.~al., Ref.~\onlinecite{tianStableElectroluminescenceAmbipolar2023}) was implemented because EL may be quenched after a few seconds of light emission. The set-reset sequence involves swapping the polarities of the topgate voltages at the rate of a few hertz. This swapping dislodges the buildup of trapped charges in the vicinity of the junction. Importantly, the set-reset sequence is compatible with high-frequency pulsing of the ohmics for the purpose of time-resolved EL (TREL) measurements\cite{tianStableElectroluminescenceAmbipolar2023} (see RF circuit in Fig.\,\ref{FigS4}).

\begin{table}[t]
    \caption{Four distinct circuit configurations for measurements on lateral 2D p\nobreakdashes--n junction. Circuit diagrams for these configurations are shown in Fig.\,\ref{FigS2}.}
    \begin{ruledtabular}
    \begin{tabular}{ccccc}
        \textbf{left side} & \textbf{\textit{V}}$_\textbf{tgL}$ & \textbf{right side} & \textbf{\textit{V}}$_\textbf{tgR}$ & \textbf{shorthand}\\
        \hline
        floating, \textit{n}-type & $>0$~V & grounded, \textit{p}-type & $<0$~V & LFN\\
        floating, \textit{p}-type & $<0$~V & grounded, \textit{n}-type & $>0$~V & LFP\\
        grounded, \textit{n}-type & $>0$~V & floating, \textit{p}-type & $<0$~V & LGN\\
        grounded, \textit{p}-type & $<0$~V & floating, \textit{n}-type & $>0$~V & LGP\\
    \end{tabular}
    \end{ruledtabular}
    \label{tab:4configs}
\end{table}

\begin{figure}[b]
    \includegraphics[width=0.9\columnwidth]{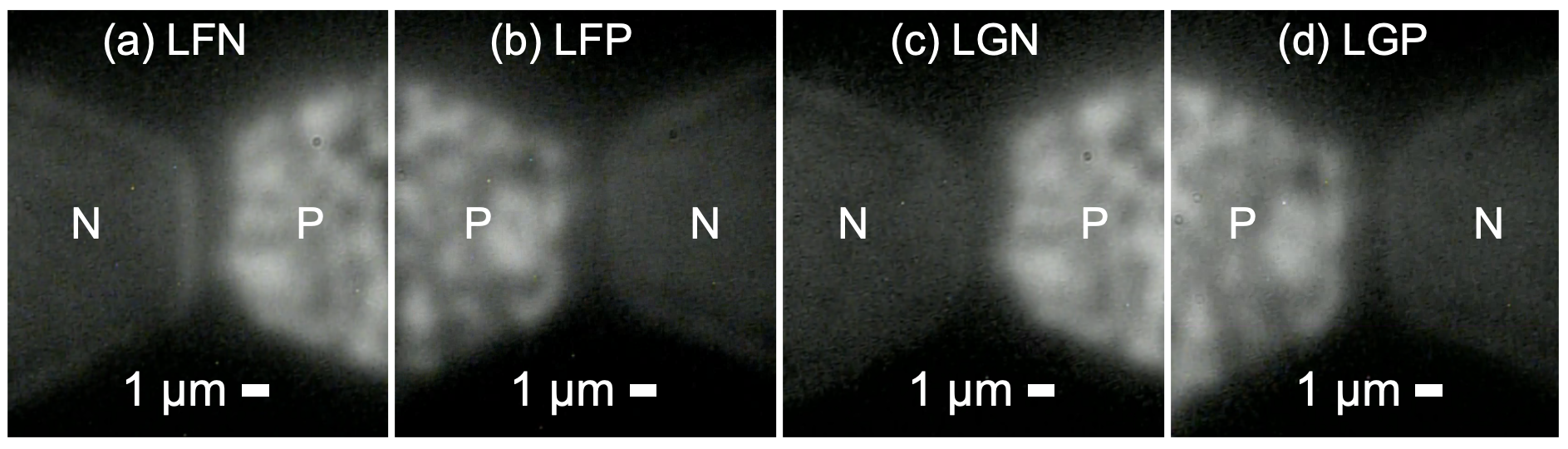}
    \caption{Still frames captured from video of the lateral p\nobreakdashes--n junction in the four measurement configurations listed in Table\,\ref{tab:4configs}. The stills are captured from the center of the sample, and clearly show the left and right trapezoidal ITO gates separated by a small gap. As mentioned in the main text, the \textit{p}-side is consistently brighter than the \textit{n}-side.}
    \label{FigS3}
\end{figure}

\begin{figure}[h]
    \includegraphics[width=0.6\columnwidth]{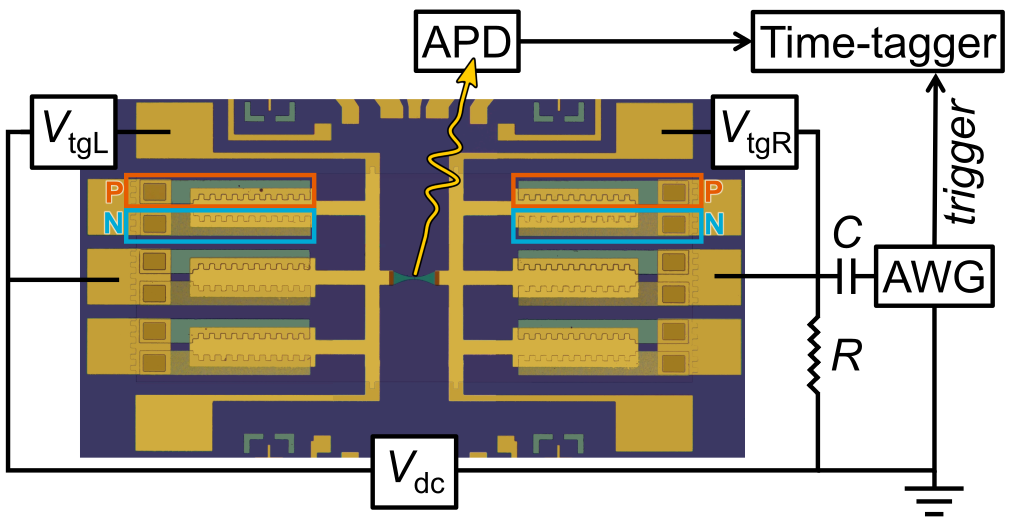}
    \caption{RF circuit for time-resolved EL measurements. The forward bias $V_\text{pn}$ comprises a subthreshold DC component $V_\text{dc}$ and an RF pulse $V_\text{rf}$ that periodically brings $V_\text{pn}$ above threshold ($V_\text{pn}$\,=\,$V_\text{dc}$\,$+$\,$V_\text{rf}$). $V_\text{dc}$ is supplied by the Keithley 2401 SourceMeter and $V_\text{rf}$ by the Tektronix 5014B AWG. Nominally, the RF waveform is a square pulse with a 10\% duty cycle and 120\,MHz frequency. In reality, the AWG has a maximum rise time of 0.95\,ns/0.6\,$V_\text{p--p}$ and the pulses are Gaussian. The \textit{start} and \textit{stop} tags for the Swabian time tagger were trigger pulses from the AWG and the APD signal, respectively.}
    \label{FigS4}
\end{figure}

Figure~\ref{FigS2} shows four possible electrical circuit configurations that may be used to operate the device. In Figure~\ref{FigS2}(a), the ohmic contacts on the \textit{right side} (R) of the device are referenced to \textit{ground} (G) and those on the \textit{left side} (L) are \textit{floating} (F), while the opposite is true for the circuit in Fig.\,\ref{FigS2}(b). Additionally, the ambipolar nature of the device allows induced charge carriers on the left side to be either \textit{n}-type (N, when $V_{\text{tgL}}>0$~V) or \textit{p}-type (P, when $V_{\text{tgL}}<0$~V), with the right side respectively being \textit{p}-type and \textit{n}-type. The resulting four distinct measurement configurations (Table~\ref{tab:4configs}) can be referred to using the shorthands LFN, LFP, LGN and LGP. This nomenclature stays valid even during set-reset operation, since it corresponds to the device configuration in the \textit{`on'} state of each set-reset cycle.\cite{tianStableElectroluminescenceAmbipolar2023} Figure~\ref{FigS3} demonstrates that observed behaviour for the $p$-side and $n$-side of the device (relative brightness, EL spectral composition and energy) was found to be independent of the specific circuit configuration, as expected.

Light was collected with an Attocube cryo-objective with a numerical aperture of 0.81 and a working distance of 0.70\,mm. EL spectral data was frequency-selected using a Princeton Instruments Acton Series SP-2750 optical spectrometer equipped with a 1200\,grooves/mm grating, and subsequently acquired with an integrated Princeton Instruments PIXIS 100 CCD camera thermoelectrically cooled to $-70$\,\textdegree{}C. Optical video of the p\nobreakdashes--n junction in all four measurement configurations was captured using a Watec WAT-910HX CCIR miniature camera with a 0.5\,inch black and white CCD; still frames from these videos are show in Fig.\,\ref{FigS3}. These images are qualitative, used only to infer the relative distribution of EL over the device.

The RF component of the source-drain bias $V_\text{pn}$ for TREL measurements was supplied by a Tektronix 5014B arbitrary waveform generator (AWG). The relevant circuit is shown in Fig.\,\ref{FigS4}. A Ti:sapphire pulsed laser was also used for time-resolved PL measurements. A Swabian Instruments Time Tagger Ultra in conjunction with an Excelitas Technologies SPCM-AQRH-16-FC avalanche photodiode (APD) were used to obtain the lifetimes of HE EL, X$^0$ EL and X$^0$ PL emission lines.

\clearpage
\newpage

\section{Transport characterization, mobility and density}
\label{sec:transport}

\begin{figure}[h]
    \centering
    \includegraphics[width=1.0\columnwidth]{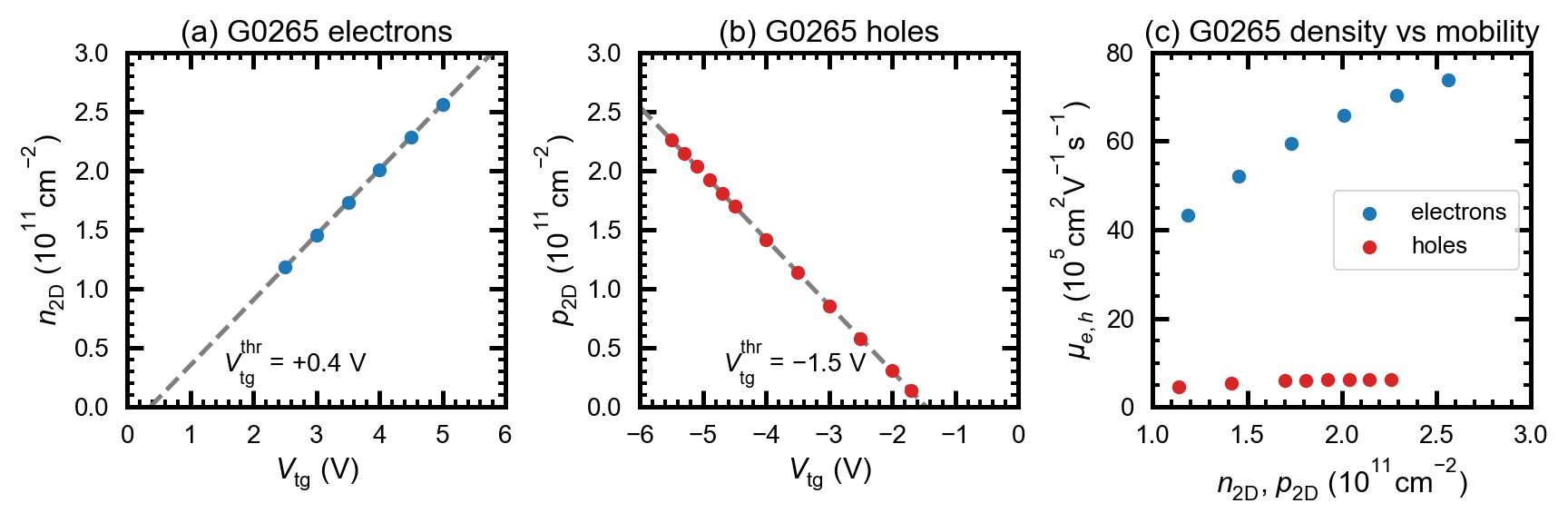}
    \caption{Four-terminal transport properties of wafer G265. (a) Electron density ($n_\textsc{2D}$) as a function of topgate voltage $V_\text{tg}$. The turn-on threshold for the 2DEG is $V_\text{tg}$ = $+0.4$\,V. (b) Hole density ($p_\textsc{2D}$) as a function of $V_\text{tg}$.  The turn-on threshold for the 2DHG is $V_\text{tg}$ = $-1.5$\,V. (c) Electron and hole mobilities ($\mu_{e,h}$) as a function of $n_\textsc{2D}$, $p_\textsc{2D}$.}
    \label{FigS5}
\end{figure}

Figure~\ref{FigS5} shows $n_\textsc{2D}(V_\text{tg})$, $p_\textsc{2D}(V_\text{tg})$, $\mu_e(n_\textsc{2D})$, and $\mu_h(p_\textsc{2D})$ in the same ambipolar Hall bar, where a 2DEG or 2DHG can induced depending on the polarity of $V_\text{tg}$. There are separate turn-on threshold voltages for the 2D carrier gas and its ohmic contacts. Any accumulation-type device only conducts when both its 2D carrier gas and ohmic contacts both turn on. In the case of the specific Hall bar shown in Fig.\,\ref{FigS5}, the $p$-type ohmic contacts turn on at a lower $V_\text{tg}$ than the 2DHG: the Hall bar starts conducting at $V_\text{tg}$\,=\,$-$1.5\,V. By contrast, the $n$-type ohmic contacts turn on at a higher $V_\text{tg}$ than the 2DEG: the Hall bar only starts conducting at $V_\text{tg}$\,$>$\,$+$2.2\,V.

\section{EL peak identification}
\label{sec:identification}

Spectral lines from bulk GaAs and GaAs/AlGaAs heterostructures have been extensively studied using PL. Peak IDs for our EL spectra were assigned using Ref.\,\onlinecite{pavesiPhotoluminescenceAlxGa1xAsAlloys1994} as the main resource.

\begin{figure}[t]
    \centering
    \includegraphics[width=0.9\columnwidth]{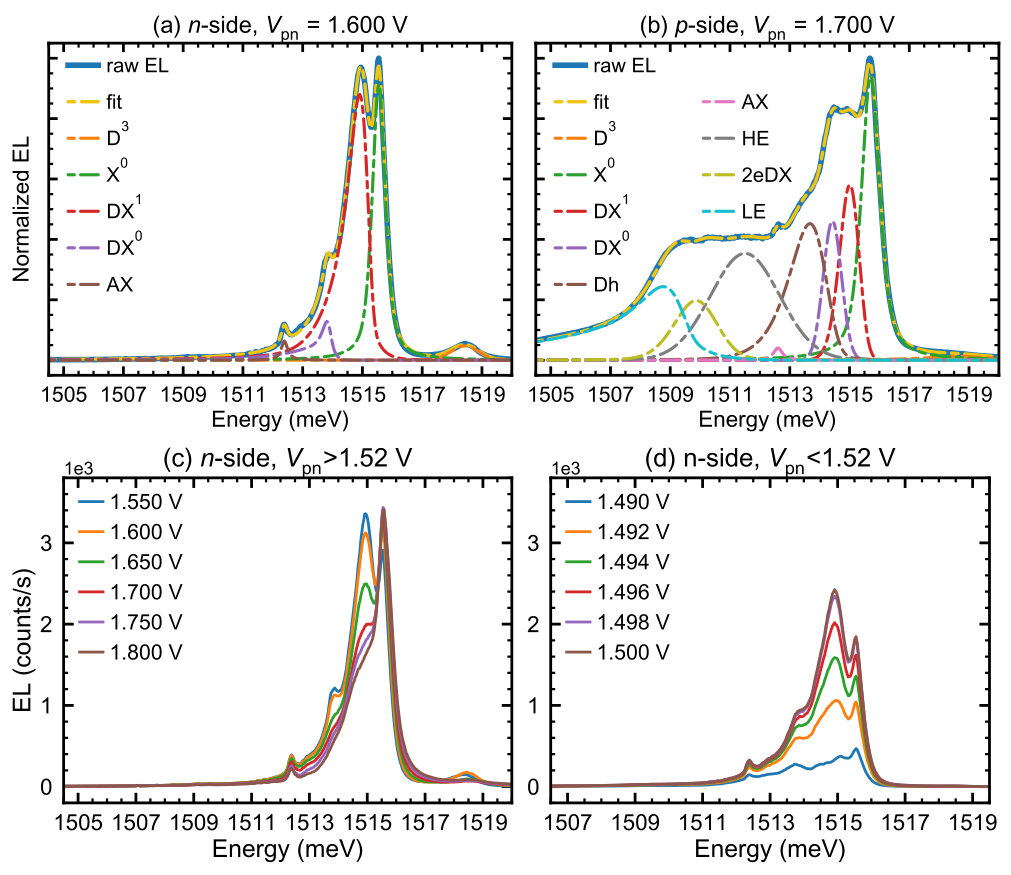}
    \caption{EL spectra from the lateral p\nobreakdashes--n junction induced at the SH. Normalized spectra with peak fits are shown for the: (a) \textit{n}-side and (b) \textit{p}-side. (c) Series of spectra on the $n$-side as a function of $V_\text{pn}$ for: (c) above-threshold operation and (d) subthreshold operation. Panels (a) and (b) are reproduced from Fig.\,2 in the main text for easy reference with Table~\ref{tab:EL_fits}; the x-axis is extended to 1519.5\,meV to show the D$^3$ peak.}
    \label{FigS6}
\end{figure}

In Figure~\ref{FigS6}(c), there is a clear correlation in intensity among three peaks (located at 1513.8\,meV, 1514.9\,meV and 1518.4\,meV) as we sweep $V_\text{pn}$, suggesting a commonality in their origin. From Ref.\,\onlinecite{pavesiPhotoluminescenceAlxGa1xAsAlloys1994}, these peaks are most likely the DX$^0$, DX$^1$ and D$^3$ transitions, respectively (see Table~\ref{tab:EL_fits} for descriptions). The remaining peaks on the \textit{n}-side are closest in energy to the AX (small peak at 1512.4\,meV) and X$^0$ (sharp peak at 1515.5\,meV) transitions.

With the knowledge gained from the \textit{n}-side spectra [Fig.\,\ref{FigS6}(a)], we can assign the same labels onto peaks with very similar energies on the \textit{p}-side. Any differences in energies are minor ($<$\,0.04\%) and may be ascribed to errors in the fitting algorithm. For $V_\text{pn}$\,$\leq$\,1.65\,V, a transition at $\sim$1517.5\,meV corresponding to the n\,=\,2 state of the neutral donor (D$^2$) is visible. Above 1.65\,V, this shifts to the D$^3$ transition at $\sim$1518.4\,meV. An interesting observation is that the AX peak on the \textit{p}-side is split into 3 distinct transitions. This has also been observed in PL and is due to spin-orbit splitting.\cite{venghausExcitationIntensityDependence1978,Stebefinestructure1981} The energy of the additional peak between AX and DX$^0$ suggests it must be the Dh transition, in which a free hole recombines with an electron bound to a neutral donor. Since the presence of free holes is required, it makes sense that the Dh peak is present only on the \textit{p}-side and not on the \textit{n}-side. Finally, the fit reveals a peak around 1509.9\,meV that lies between the LE and HE peaks of the H\nobreakdash-band. Unlike the H-band, this peak does not increase in energy with increasing $V_\text{pn}$ during subthreshold operation. Comparing this with the peak assignments in Ref.\,\onlinecite{pavesiPhotoluminescenceAlxGa1xAsAlloys1994}, we think this is the 2eDX transition, or the two-electron transition of DX$^0$ with the donor left in the n\,=\,2 state.

Figures~\ref{FigS6}(c) and \ref{FigS6}(d) show changes in EL spectra as a function of $V_\text{pn}$. The behavior is quite similar to that observed for the $p$-side, described in the main text. For above-threshold operation ($V_\text{pn} > 1.52$~eV), the impurity peaks decrease in intensity in favor of X$^0$, which increases with increasing $V_\text{pn}$. For sub-threshold operation ($V_\text{pn} < 1.52$~eV), the impurity peaks increase in intensity with increasing $V_\text{pn}$.

\begin{table}[t]
    \caption{EL peak fit parameters for emission lines from the \textit{p}-side and \textit{n}-side, along with descriptions. The primary source for these peak IDs (besides the H-band) is Ref.\,\onlinecite{pavesiPhotoluminescenceAlxGa1xAsAlloys1994}.}
    \begin{ruledtabular}
    \begin{tabular}{cccl}
        peak ID & $E_\text{EL}$\,(p) (meV) & $E_\text{EL}$\,(n) (meV) & peak description\\
        \hline
        D$^3$ & 1518.7 & 1518.4 & n\,=\,3 state of neutral donor\\
        X$^0$ & 1515.7 & 1515.5 & n\,=\,1 state of neutral free exciton\\
        DX$^1$ & 1515.0 & 1514.9 & excited state of DX$^0$\\
        DX$^0$ & 1514.4 & 1513.8 & exciton bound to neutral donor\\
        Dh & 1513.6 & --- & free hole\,+\,donor-bound electron\\
        AX & 1512.6 & 1512.4 & exciton bound to neutral acceptor\\
        HE & 1511.6 & --- & H-band high-energy peak\\
        2eDX & 1509.9 & --- & two-electron transition of DX$^0$\\
        LE & 1508.8 & --- & H-band low-energy peak\\
    \end{tabular}
    \end{ruledtabular}
    \label{tab:EL_fits}
\end{table}

\clearpage
\newpage

\section{H-band and X$^0$ temperature dependence}
\label{sec:Tdep}

\begin{figure}[b]
    \includegraphics[width=0.50\columnwidth]{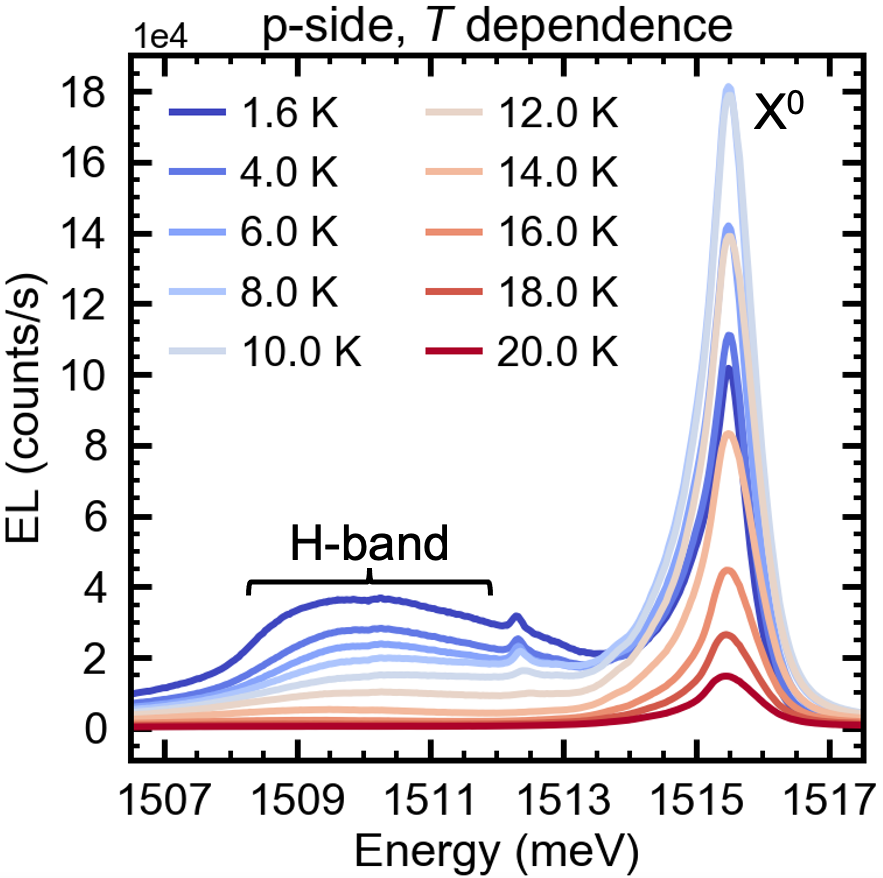}
    \caption{Temperature dependence of EL spectra from the \textit{p}-side of the device for $V_\text{pn}$\,=\,1.7\,V. For the sake of comparison, EL counts for each spectrum are scaled with respect to their corresponding acquisition times. }
    \label{FigS7}
\end{figure}

Figure~\ref{FigS7} shows the temperature dependence of X$^0$ and H\nobreakdash-band. The intensity of the H-band's HE and LE (located at energies below 1512\,meV) monotonically decrease as temperature increases. HE and LE completely disappear when $T$\,$>$\,16\,K. The intensity of the X$^0$ does not monotonically decrease as temperature increases, and X$^0$ completely disappears when $T$\,$>$\,24\,K. The higher operating temperature of X$^0$ relative to HE indicates X$^0$ has higher binding and bound state energies than HE, a true 2D free exciton. This is consistent with X$^0$'s higher EL energy than PL energy, and reinforces the idea that the EL X$^0$ from our lateral p\nobreakdash-n junctions is a true 2D exciton. In this context, the non-monotonic behavior of X$^0$ with temperature is easy to explain: the intensities of HE and LE decrease as a function of temperature in favor of the higher energy transition X$^0$ from 1.6\,K to 10\,K (similar behavior is observed when increasing $V_\text{pn}$). Above 10\,K, the X$^0$ intensity starts decreasing with increasing temperature. Overall, X$^0$ in SH heterostructures is much less robust against temperature (extinction when $T$\,$>$\,24\,K) than in QW heterostructures (extinction when $T$\,$>$\,85\,K), as expected from an exciton with much weaker confinement.

\clearpage
\newpage

\section{H-band time-resolved EL}
\label{sec:TREL-Hband}

\begin{figure}[h]
    \centering
    \includegraphics[width=0.49\columnwidth]{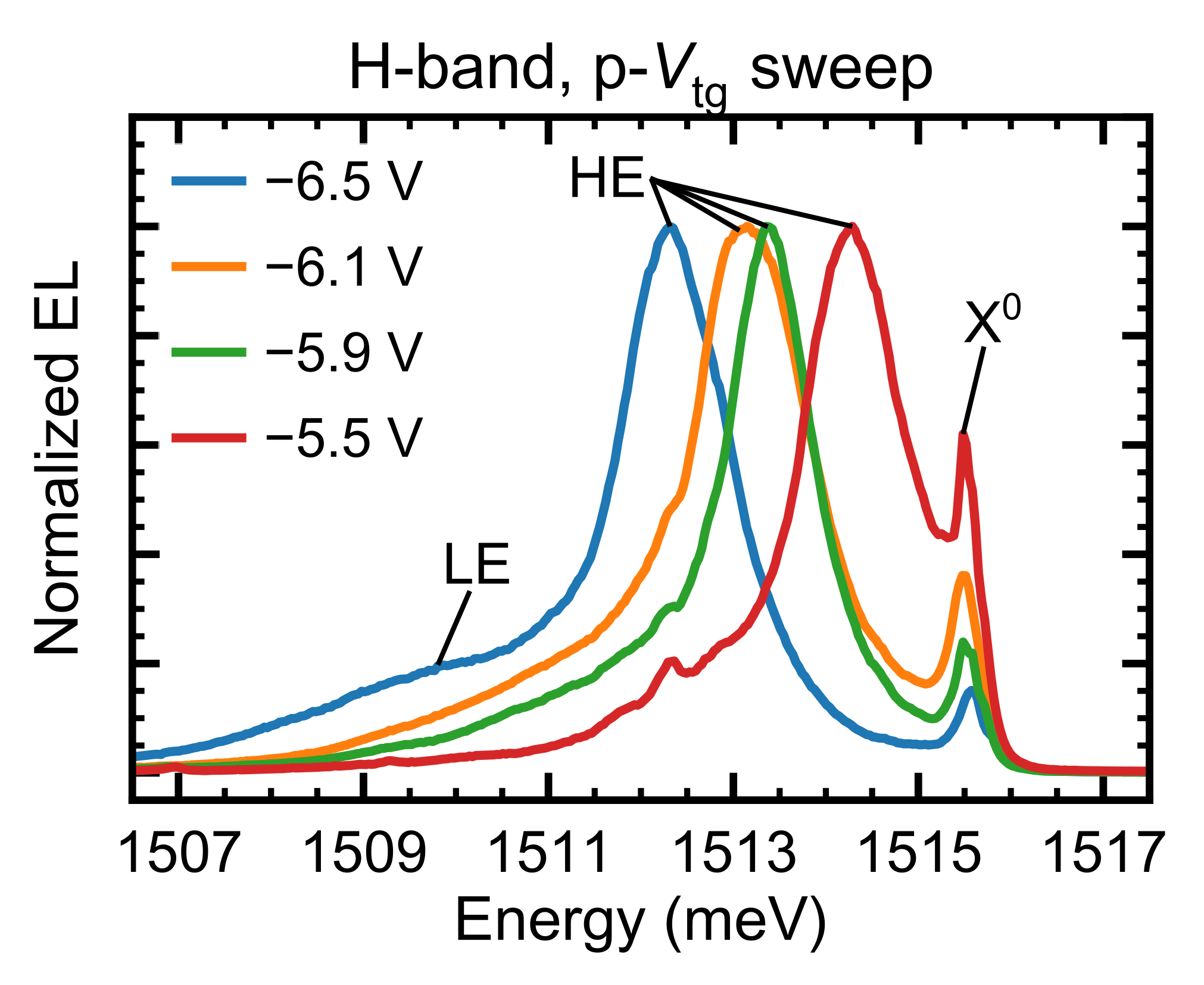}
    \caption{EL spectra with HE tuned to different energies by changing the \textit{p}-side $V_\text{tg}$. The intensity of LE and other impurity-related excitons are minimized by collecting EL from the $p$-side, away from the gap between left/right topgates (i.e., far away from the  center of the p-n junction). The HE energy increases as $|V_\text{tg}|$ decreases. In contrast, the X$^0$ energy remains fixed at $\sim$1515.5\,meV for all $V_\text{tg}$ values.}
    \label{FigS8}
\end{figure}

\begin{table}[h]
    \begin{ruledtabular}
    \begin{tabular}{cccccc}
        $V_\text{dc}$\,(V) & $V_\text{rf}$\,(V$_\text{p--p}$) & $V_\text{pn}$\,(V) & $V_\text{tg}$\,(V) & $E_\text{EL}$(HE)\,(meV) & $\tau_\text{EL}$(HE)\,(ps)\\
        \hline
        1.492 & 0.113 & 1.605 & $-5.5$ & 1514.3 & 575\,$\pm$\,6\\
        1.498 & 0.119 & 1.617 & $-5.9$ & 1513.4 & 672\,$\pm$\,4\\
        1.500 & 0.144 & 1.644 & $-6.1$ & 1513.2 & 714\,$\pm$\,7\\
        1.429 & 0.452 & 1.881 & $-6.5$ & 1512.3 & 747\,$\pm$\,7
    \end{tabular}
    \end{ruledtabular}
    \caption{HE energies and lifetimes for different \textit{p}-side topgate voltages $V_\text{tg}$. For all spectra in Fig.\,\ref{FigS8}, the forward bias $V_\text{pn}$\,=\,$V_\text{dc}$\,$+$\,$V_\text{rf}$ (see circuit in Fig.\,\ref{FigS2}) was used to increase photon counts, and was above threshold in all cases ($V_\text{pn}$\,$>$\,1.52\,V). As per Fig.\,2(d) in the main text, the H-band energy is not affected by changing $V_\text{pn}$ for above-threshold operation. The tuning of $E_\text{EL}$(HE) is thus only accomplished by changing $V_\text{tg}$. }
    \label{tab:TREL}
\end{table}

\clearpage
\newpage

\begin{table}[b]
    \begin{ruledtabular}
    \begin{tabular}{lcccccc}
        wafer & heterostructure & $\tau_\textsc{pl}$(X$^0$) & $\tau_\textsc{el}$(X$^0$) &  $\tau_\textsc{pl}/\tau_\textsc{el}$ & $E_\textsc{pl}$(X$^0$) & $E_\textsc{el}$(X$^0$) \vspace{-2.5 mm} \\
        ~~ID & type & (ps) & (ps) & -- & (meV) & (meV) \\
        \hline
        G370 & SH & 1300 & -- & -- & 1515.5 & -- \\
        G265 & SH & 1610 & 337 & 4.8 & 1515.3 & 1515.7 \\  
        G375 & QW & 419 & 237 & 1.8 & 1534.1 & 1534.4 \\   
        G569 & QW & 400 & 231 & 1.7 & 1534.3 & 1534.7      
    \end{tabular}
    \end{ruledtabular}
    \caption{Exciton lifetimes and energies of X$^0$ for PL/EL in four wafers. G375 and G569 have a 15 nm wide QW. The $E_\textsc{el}$ from G265 and G375 were from the $p$-side and $n$-side, respectively. }
    \label{tab:TREL}
\end{table}

\section{X$^\textbf{0}$ time-resolved PL and EL}
\label{sec:TRELPL-X0}

\begin{figure}[h]
    \includegraphics[width=0.90\columnwidth]{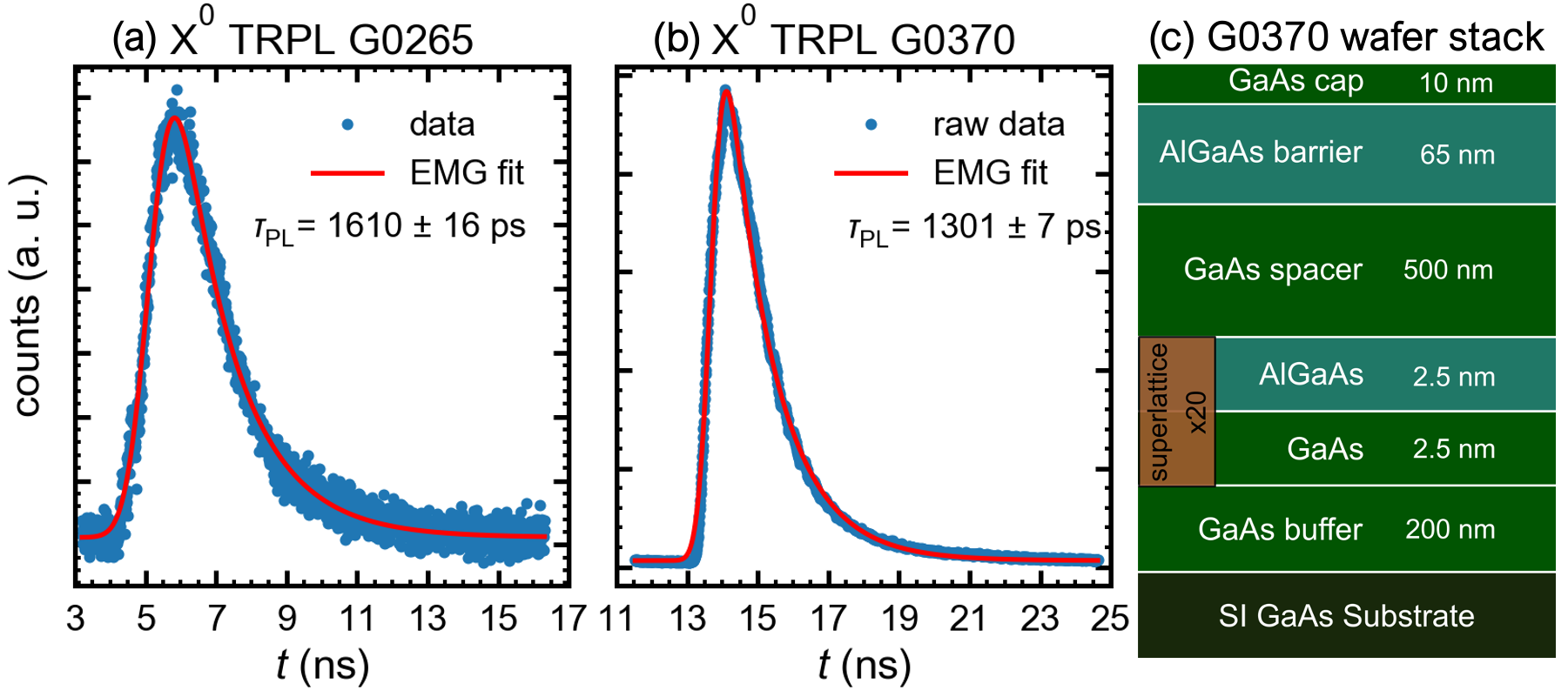}
    \caption{(a) Time-resolved photoluminescence (TRPL) of X$^0$ from wafer G265, which was the base for all devices studied in the main text. (b) TRPL of X$^0$ for wafer G370, a similar GaAs/AlGaAs heterostructure wafer with a 75\,nm deep SH. (c) Wafer stack for G370.}
    \label{FigS9}
\end{figure}

\begin{figure}[h]
    \includegraphics[width=0.95\columnwidth]{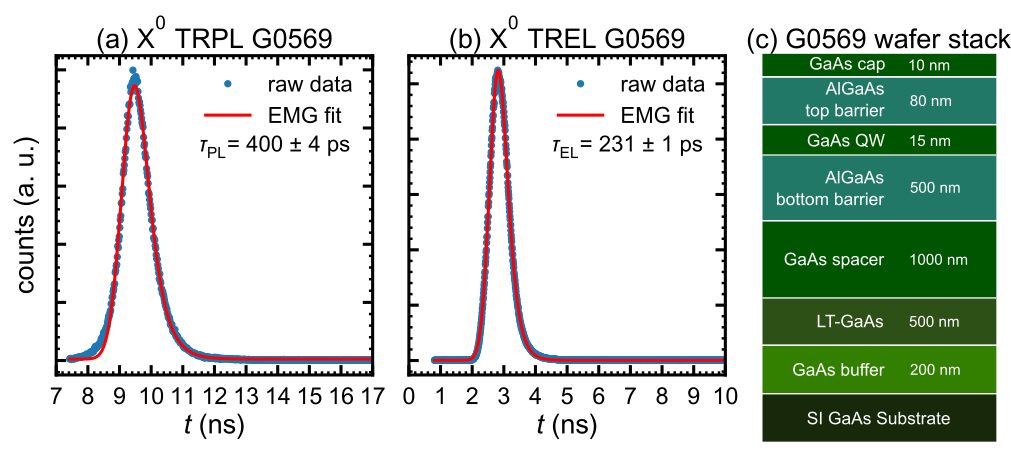}
    \caption{(a) Time-resolved photoluminescence (TRPL) of X$^0$ from wafer G569. (b) Time-resolved electroluminescence (TREL) of X$^0$ for wafer G569. (c) Wafer stack for G569.}
    \label{FigS10}
\end{figure}

Table~\ref{tab:TREL} summarizes lifetimes and energies of X$^0$ from PL/EL in four SH/QW heterostructures. Wafers G370 and G375 were grown the same week. Wafer G265 was grown three months earlier, within the same MBE campaign. Wafer G569 was grown 17 months after G370/G375. Figures~\ref{FigS9} and \ref{FigS10} show time-resolved PL and EL data analysis for some of the heterostructures listed in Table~\ref{tab:TREL}.

The data in Table~\ref{tab:TREL} from all wafers are very consistent with each other. The $\tau_\textsc{pl}$ from the two SH heterostructures are similar to each other (1300\,ps, 1610\,ps), and the $\tau_\textsc{pl}$ from the two QW heterostructures are similar to each other (400\,ps, 419\,ps). Even among the two SH heterostructures, $\tau_\textsc{pl}$ scales with electron mobility, with $7 \times 10^6$~cm$^2$/Vs for G265 and $2 \times 10^6$~cm$^2$/Vs for G370 at the same 2DEG density. Finally, all the $E_\textsc{el}$ are larger than their corresponding $E_\textsc{pl}$ from the same wafer.

Three observations from Table~\ref{tab:TREL} support our claim that EL X$^0$ from lateral p\nobreakdash-n junctions have a 2D excitonic origin. First, the $\tau_\textsc{pl}$ for SH heterostructures are much larger than the ones for QW heterostructures, partially on account of their larger mobilities but also on account of their higher dimensional excitonic origin: 3D (bulk) for SH heterostructures whereas 2D for QW heterostructures. Second, all the $\tau_\textsc{el}$ from both SH and QW heterostructures are clustered together (231$-$337\,ps), possibly denoting a similar dimensional origin (2D). Third, the two QW heterostructures yield nearly identical ratios $\tau_\textsc{pl}/\tau_\textsc{el}\approx 1.7, 1.8$ which are much smaller than the ratio $\tau_\textsc{pl}/\tau_\textsc{el}=4.8$ from the SH heterostructure.

When performing TRPL X$^0$ in a SH heterostructure, care must be taken not to collect X$^0$ photons from the SI GaAs substrate or the LT GaAs layer, which have much shorter lifetimes (on the scale of 300$-$500 ps). The quality of the GaAs material can profoundly affect its optical properties. The long $\tau_\textsc{pl}$ and high mobility values emphasizes the low disorder/scattering from the MBE-grown bulk GaAs in our SH heterostructures.

\clearpage
\newpage

\begin{figure}[t]
    \includegraphics[width=1.0\columnwidth]{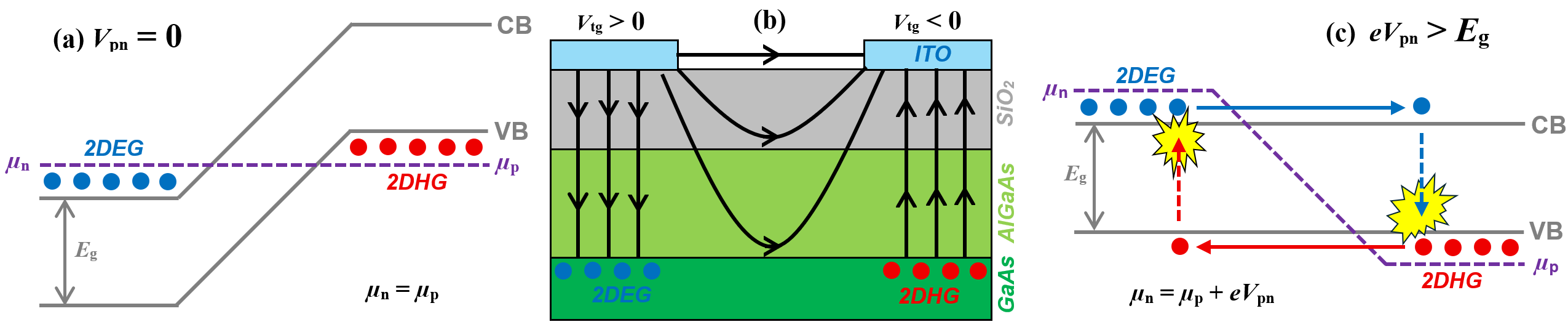}
    \caption{Lateral p\nobreakdashes--n junction in a SH heterostructure. (a) Band structure diagram at equilibrium ($V_\text{pn}$\,=\,0), along the 2DEG/2DHG plane. Both 2DEG and 2DHG are turned on ($|V_\text{tg}|$\,=\,5\,V) and have high carrier densities. (b) Cross-sectional view of the corresponding electric field distribution diagram at equilibrium ($V_\text{pn}$\,=\,0). The diagram is not to scale. (c) Band structure diagram at non-equilibrium ($eV_\text{pn}$\,$>$\,$E_\text{g}$), along the plane of the GaAs/AlGaAs interface. Symbols represent: the conduction band (CB), the valence band (VB), the GaAs band gap ($E_\text{g}$), the chemical potential levels on the $n$-side ($\mu_\text{n}$) and $p$-side ($\mu_\text{p}$), and the forward bias ($V_\text{pn}$).}
    \label{Fig:Efields}
\end{figure}

\section{Electric field profiles and transit times}
\label{sec:discussion}

Figure~\ref{Fig:Efields}(a) shows the band structure of a lateral p\nobreakdashes--n junction at equilibrium ($V_\text{pn}=0$) in a SH heterostructure, as a function of position in the plane of the 2DEG/2DHG, along the GaAs/AlGaAs interface. Almost all of the out-of-plane electric field ($E_\perp$) will be generated by the topgates from the $p$-side and $n$-side, and any forward bias would only contribute to the in-plane electric field ($E_\parallel$). Figure~\ref{Fig:Efields}(b) shows a diagram of the electric field distribution associated with Fig.\,\ref{Fig:Efields}(a). Crucially, at the center of the gap between the topgates, the out-of-plane electric field contribution from the topgates will be zero ($E_\perp=0$). A small non-zero $E_\perp$ component might be present between the center of the gap and the edge of the 2DEG, with an identically small component of opposite sign existing on the 2DHG side. Almost all of the in-plane electric field driving diode current at non-equilibrium originates from the forward bias ($E_\parallel$~$\approx$~$V_\text{pn}$/gap~=~1.5\,V/5\,\textmu m~$=$~3\,kV/cm), as depicted in Fig.\,\ref{Fig:Efields}(c). Thus, most carriers (electrons or holes) entering and being accelerated across the topgate gap will not stray far from the GaAs/AlGaAs interface. Carriers originating from the 2DEG (2DHG) will re-enter the $p$-side ($n$-side) directly into the plane of the 2DHG (2DEG).

Once across the gap, minority carriers are surrounded by outnumbering majority carriers. Minority carriers are severely slowed down (from their acceleration across the gap) from the strongly attractive Coulomb interactions between minority/majority carriers. Furthermore, the surrounding majority carriers provide intra-layer screening from the topgate to the minority carriers, momentarily preventing their expulsion into the substrate. Both mechanisms together give enough dwell time to the minority carriers to recombine radiatively and non-radiatively with majority carriers. EL is generated under the topgates, near the p\nobreakdashes--n junction.

Table \ref{tab:transit} shows the calculated transit times of carriers across the 5 \textmu m gap of the lateral p\nobreakdashes--n junction. For electrons, we conservatively used the bulk GaAs mobility $\mu$ found from modeling the frequency response of the tidal effect in our devices (see main text) to obtain the drift velocity $v_d = \mu \cdot E_\parallel$. The resulting transit time ($t_\text{max} \approx 72$~ps) represents an upper estimate. In the presence of quasi-ballistic transport across the gap (or low levels of small-angle scattering), the drift velocity can become much larger, up to a maximum value set by the Fermi velocity $v_\textsc{f} = \hbar \sqrt{2\pi n_\textsc{2d}}/m^*$ (where $m^*$ is the effective mass and $n_\textsc{2d}$ is the carrier density) attained in the nearby 2DEG. The resulting transit time ($t_\text{min} \approx 23$~ps) thus represents a lower estimate. The real electron transit time is most likely a value in-between the upper/lower estimates. Corresponding calculations for holes are also shown. Interaction time between electrons and holes will be determined by the carrier with the shortest transit time: the electrons.\\

\begin{table}[h]
    \begin{ruledtabular}
    \begin{tabular}{lcccccc}
         carrier & mobility & $E_\parallel$ & $v_\text{d}$ & $t_\text{max}$ & $v_\textsc{f}$ & $t_\text{min}$ \vspace{-2.5 mm} \\
        ~type & (cm$^2$/Vs) & (V/cm) & (cm/s) & (ps) & (cm/s) & (ps) \\
        \hline
        electrons & 2.3$\times$10$^3$ & 3.0$\times$10$^3$ & 6.9$\times$10$^6$ & 72 & 2.2$\times$10$^7$ & 23 \\
        holes & 4.4$\times$10$^2$ & 3.0$\times$10$^3$ & 1.3$\times$10$^6$ & 380 & 3.6$\times$10$^6$ & 140
    \end{tabular}
    \end{ruledtabular}
    \caption{Calculated transit times (min/max) for carriers to cross the gap between the $p$-side and $n$-side topgates, using $t_\text{max} = 5$~\textmu m/$v_\text{d}$ and $t_\text{min} = 5$~\textmu m/$v_\textsc{f}$. }
    \label{tab:transit}
\end{table}

\newpage